\documentclass{ws-procs9x6}

\begin{document}

\title{NEUTRON AND PROTON TESTS OF DIFFERENT TECHNOLOGIES FOR THE
  UPGRADE OF COLD READOUT ELECTRONICS OF THE ATLAS HADRONIC ENDCAP CALORIMETER}

\author{M. NAGEL}

\address{Max-Planck-Institut f\"{u}r Physik,\\
F\"{o}hringer Ring 6, 80805 Munich, Germany\\
E-mail: nagel@mppmu.mpg.de\\
www.mppmu.mpg.de}

\author{on behalf of the HECPAS Collaboration\\
(IEP Ko\v{s}ice, Univ. of Montr\'{e}al, MPI Munich, IEAP Prague, NPI \v{R}e\v{z})}

\begin{abstract}
The expected increase of total integrated luminosity by a factor of
ten at the HL-LHC compared to the design goals for LHC essentially
eliminates the safety factor for radiation hardness realized at the
current cold amplifiers of the ATLAS Hadronic Endcap Calorimeter
(HEC). New more radiation hard technologies have been studied: SiGe
bipolar, Si CMOS FET and GaAs FET transistors have been irradiated
with neutrons up to an integrated fluence of $2.2 \cdot 10^{16} \,
\mathrm{n}/ \rm{cm^{2}}$ and with 200 MeV protons up to an integrated
fluence of $2.6 \cdot 10^{14} \, \rm{p}/ \rm{cm}^{2}$. Comparisons of transistor parameters such as the gain for both types of irradiations are presented.
\end{abstract}

\keywords{ATLAS, Liquid Argon calorimeter, HL-LHC, cold readout electronics,
  radiation hardness}

\bodymatter

\section{The ATLAS Hadronic Endcap Calorimeter}

The hadronic endcap calorimeter (HEC) of the ATLAS experiment\cite{ATLAS}
at the CERN Large Hadron Collider (LHC) is a copper-liquid argon
sampling calorimeter in a flat plate design\cite{LArTDR,MPP-2007-237}. The
calorimeter provides coverage for hadronic showers in the
pseudorapidity range $1.5 < |\eta| < 3.2$. The HEC shares each of the
two liquid argon endcap cryostats with the electromagnetic endcap
(EMEC) and forward (FCAL) calorimeters, and consists of two wheels per
endcap.

\begin{figure}\label{PSB}
  \begin{center}
    \includegraphics[scale=0.38]{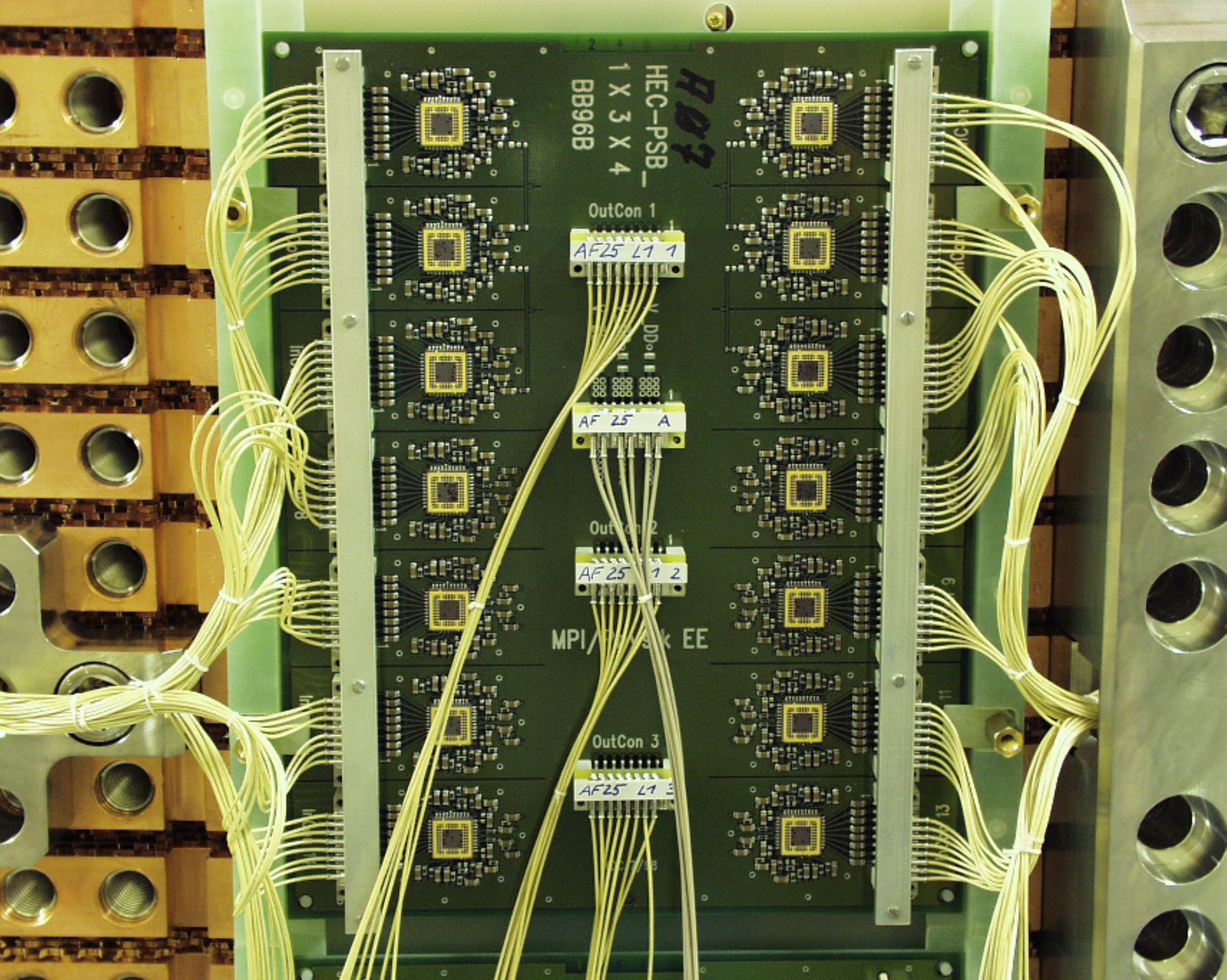}
    \caption{A detailed view of a PSB board mounted on the perimeter of a HEC wheel.}
  \end{center}
\end{figure}

A HEC wheel is made of 32 modules, each with 40 liquid argon
gaps, which are instrumented with active read-out pads. The signals
from the read-out pads are sent through short coaxial cables to
preamplifier and summing boards (PSB) mounted on the perimeter of the
wheels inside the liquid argon cryostat. The PSB boards carry highly-integrated
preamplifier and summing amplifier chips in Gallium-Arsenide (GaAs)
technology. The signals from a set of preamplifiers are summed to one
output signal, which is transmitted to the cryostat
feed-through\cite{MPP-2005-193}. Figure \ref{PSB} shows a PSB board
mounted and connected on the perimeter of a HEC wheel.


\section{Requirements of the HEC cold electronics for the HL-LHC
  upgrade}

The GaAs technology currently employed in the HEC cold electronics has
been selected for its excellent high frequency performance, stable
operation at cryogenic temperatures, and radiation hardness. The
radiation hardness specifications were defined for ten years of
operation at the LHC design luminosity of $10^{34} \, \rm{cm^{-2} \, s^{-1}}$,
including a safety factor of ten. For the high-luminosity upgrade
of the LHC (HL-LHC), the luminosity is foreseen to increase by a
factor of 5--10, effectively eliminating the safety factor. The ATLAS collaboration
therefore decided to re-examine the radiation hardness of the current
HEC cold electronics and of potential alternative
technologies\cite{Schacht}. Detailed studies of the expected radiation levels after
ten years of running under HL-LHC conditions yielded the
following requirements (including a safety factor of 10) for the HEC cold electronics:

\begin{itemlist}
\item Neutron fluence of $2 \cdot 10^{15} \, \mathrm{n} / \rm{cm}^{2}$
\item Proton fluence of $2 \cdot 10^{14} \, \mathrm{p} / \rm{cm^{2}}$
\item Gamma dose of 20 kGy
\end{itemlist}

\section{Tests}

The neutron irradiation tests were performed at the \v{R}e\v{z} Neutron
Physics Laboratory near Prague in the Czech Republic, up to an
integrated fluence of $2.2 \cdot 10^{16} \, \mathrm{n} /
\rm{cm}^{2}$. A $37\, \rm{MeV}$ proton beam incident on a $\rm{D_2O}$
target created a somewhat divergent beam of neutrons with a flux
density up to $10^{11} \, \mathrm{n} / \rm{cm^{2}} / \rm{s}$. The
proton irradiation test were performed at the Proton Irradiation
Facility at the Paul-Scherrer-Institut in Switzerland with a $200\,
\rm{MeV}$ proton beam up to an integrated fluence of $2.6 \cdot
10^{14} \, \mathrm{p} / \rm{cm}^{2}$. The transistors were bonded in 
ceramic casings and mounted on boards, which were then aligned in the 
particle beams. The three different transistor technologies being tested 
were Si CMOS FET, SiGe Bipolar HBT (both SGB25V 250nm from IHP), and
 Triquint CFH800 250nm GaAs FET pHEMT, respectively. The performance of 
the transistors was monitored during irradiation with a
vector network analyzer recording a full set of $S$-parameters. Beam
current measurements and radiation films placed at various distances
along the beam were used to determine the particle flux and the beam
profile. 

\section{Results}

\begin{figure}[b]\label{Result_figure}
  \begin{center}
    \includegraphics[scale=0.6]{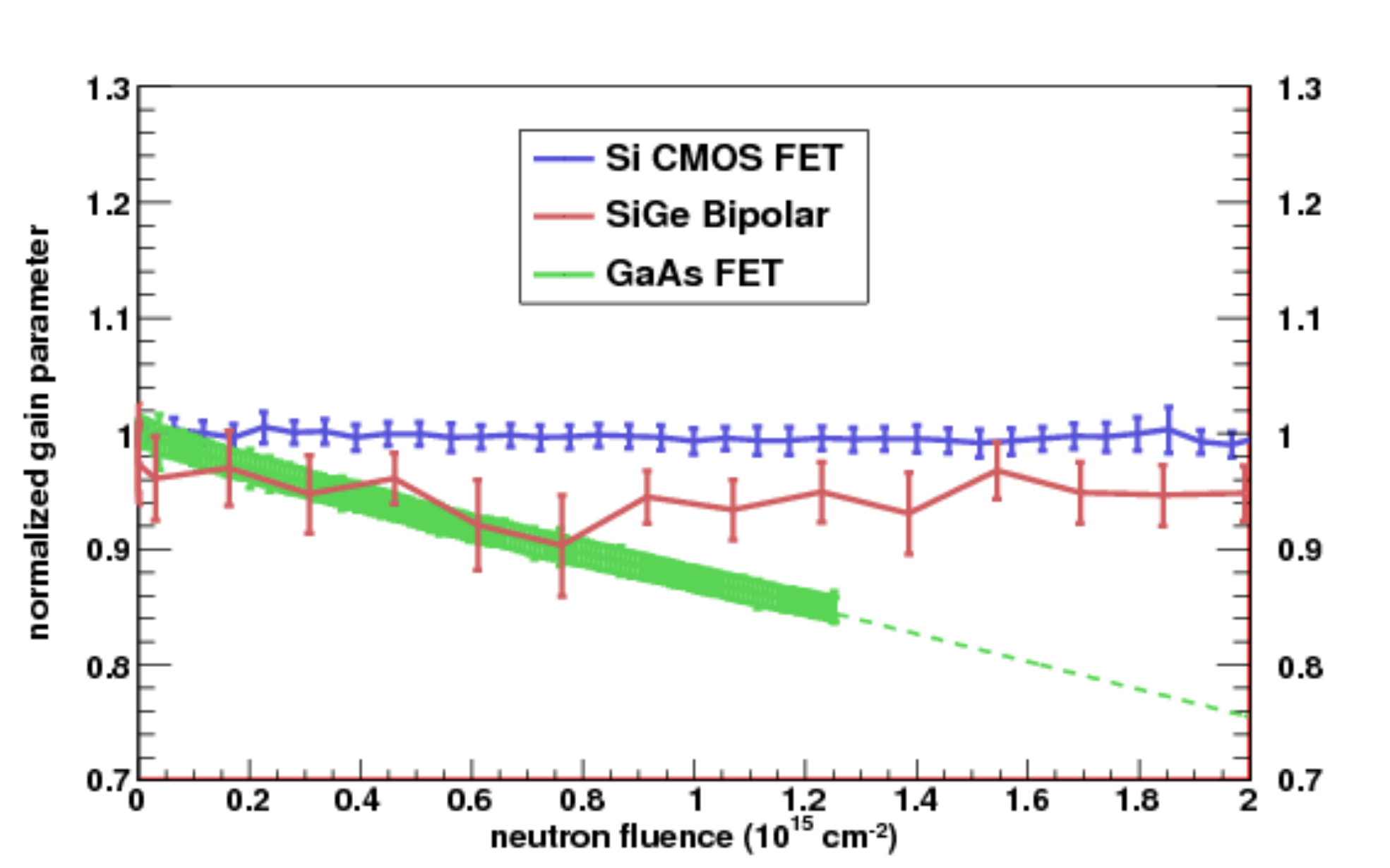}\\
    \includegraphics[scale=0.6]{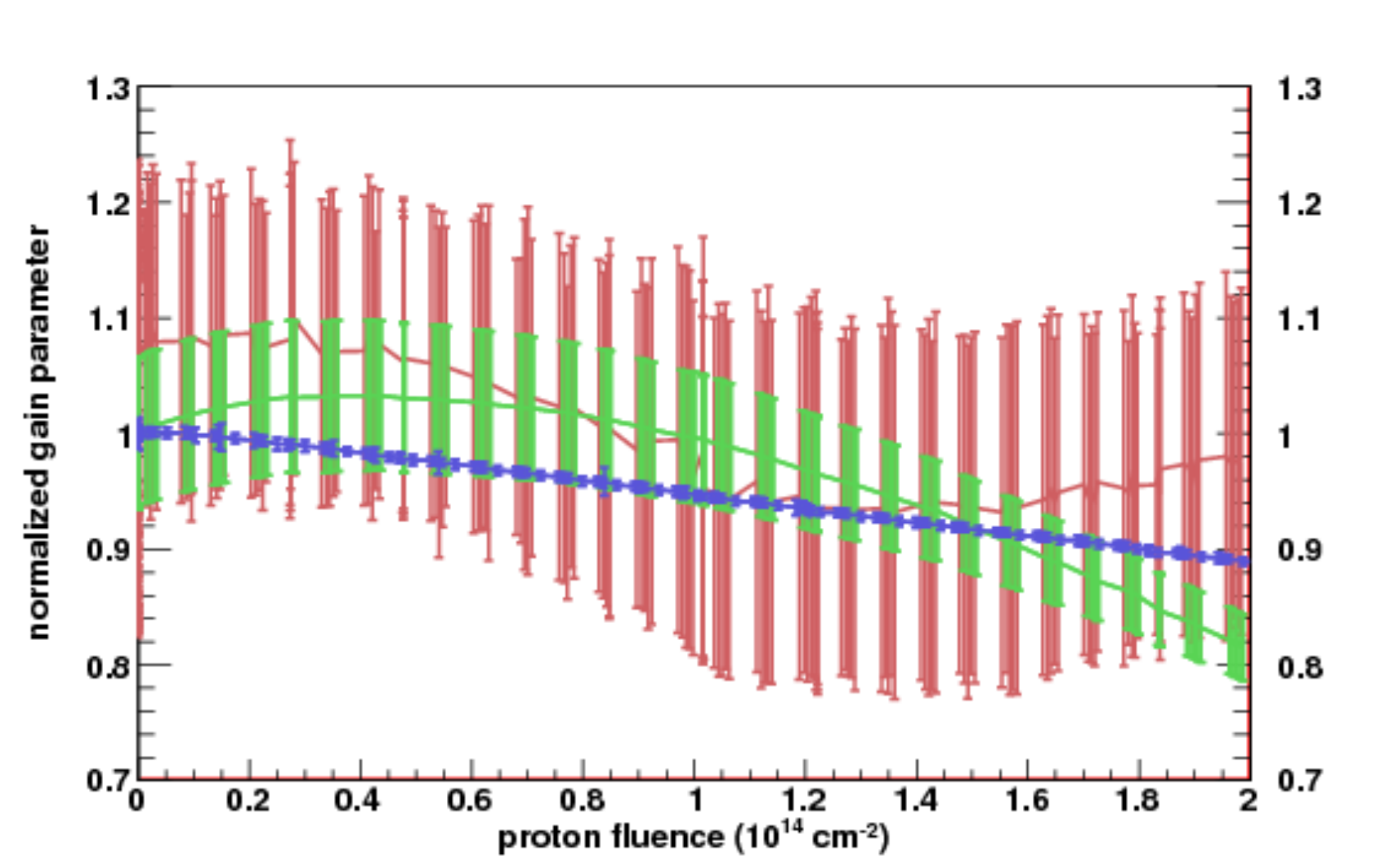}
    \caption{Relative gain loss as a function of neutron (top) and proton
      (bottom) fluence for various technologies.}
  \end{center}
\end{figure}

The various transistor parameters were calculated from the measured
$S$-parameters using standard small signal circuit models. The
transistor parameters were averaged over a certain frequency range to
obtain their mean and rms-values for every set of $S$-parameters. This
frequency range extended from $300\, \rm{kHz}$ to $100\, \rm{MHz}$,
unless a certain parameter was unstable at low or high 
frequencies, in which case appropriate cuts were applied. These averaged
transistor values were then used to characterize their behaviour as a
function of radiation.  

Figure 2 summarizes the results for the neutron (top) and the
proton (bottom) tests in terms of the appropriate gain
parameters. Displayed are the various gain parameters, i.e. the real 
part of the transconductance $g_m$ for the FET transistors and the 
current gain $\beta$ for the Bipolar transistors, as a function of 
the corresponding particle fluence up to the required limit, normalized 
to the corresponding value before irradiation. We apply a linear fit 
to the observed radiation dependence, and the relative change of the 
gain parameter resulting from the fit is quoted in Table 1.

\begin{table}\label{Result_table}
  \tbl{Loss of gain of various technologies under neutron and
    proton irradiation.}
  {\begin{tabular}{lcc}\toprule
    \multicolumn{1}{c}{Technology} & Neutron fluence & Proton fluence\\
    & $2 \cdot 10^{15} \, \mathrm{n} / \rm{cm}^{2}$ & $2 \cdot 10^{14} \,
    \mathrm{p} / \rm{cm}^{2}$\\
    \colrule
    Si CMOS FET & -1 \% & -11 \%\\
    SiGe Bipolar & -3 \% & -10 \%\\
    GaAs FET & -24 \% & -21 \%\\
    \botrule
  \end{tabular}}
\end{table}

The results show that the alternative technologies under consideration
are more radiation hard than the current GaAs technology, in
particular for neutron irradiation. However, the loss in gain for the
GaAs FETs is only moderate, and could be considered acceptable,
especially if occuring uniform across the HEC. Gamma irradiation tests
are planned for the near future to complete the radiation hardness tests.

\bibliographystyle{ws-procs9x6}
\bibliography{icatpp2011}

\end{document}